\documentclass{ws-procs975x65}

\newcommand{\abs}[1]{\left|#1\right|}
\newcommand{\norm}[1]{\left|\!\left|#1\right|\!\right|}
\newcommand{\dom}[1]{\ensuremath{\mathcal{D}_{#1}}}
\newcommand{\opr}[1]{\ensuremath{\mathbf{\mathsf{#1}}}}
\newcommand{\mcal}{\ensuremath{\mathcal}}

\newcommand{\braket}[2]{\ensuremath{\left<\left.#1\right|#2\right>}}
\newcommand{\ketbra}[2]{\ensuremath{\left|#1\left>\right<#2\right|}}

\newcommand{\expo}[1]{\ensuremath{\mbox{e}^{#1}}}

\newcommand{\kernel}[1]{\ensuremath{\left.\left<q\right|#1\left|q'\right>\right.}}

\newcommand{\oprD}[1]{\ensuremath{#1:\mathcal{D}_{#1}\subseteq \mathcal{H}\mapsto \mathcal{H}}}

\begin{document}

\title{What could have we been missing while Pauli's Theorem was in force?\thanks{To appear in the proceedings of the ``International Colloquium in Time and Matter,'' Venice, Italy, August 11-24, 2002 (World Scientific).}}

\author{Eric A. Galapon}

\address{Theoretical Physics Group, National Institute of Physics\\University of the Philippines, Diliman, Quezon City, 1101 Philippines\\E-mail: egalapon@nip.upd.edu.ph}


\maketitle

\abstracts{Pauli's theorem asserts that the canonical commutation relation $[T,H]=iI$ only admits Hilbert space solutions that form a system of imprimitivities on the real line, so that only non-self-adjoint time operators exist in single Hilbert quantum mechanics. This, however, is contrary to the fact that there is a large class of solutions to $[T,H]=iI$, including self-adjoint time operator solutions for semibounded and discrete Hamiltonians. Consequently the theorem has brushed aside and downplayed the rest of the solution set of the time-energy canonical commutation relation.}

\section{Introduction}\label{intro}

Time is widely recognized as a parameter in quantum mechanics, and no one emphatically asserts this conviction in recent times more than Sakurai's assertion in his well known textbook---``The first important point we should keep in mind is that time is just a parameter in quantum mechanics, not an operator. In particular, time is not an observable. It is nonsensical to talk about the time operator in the same sense as we talk about the position operator.''\cite{sakurai} On the other hand, the scalar status of time has been seen as a weakness of quantum theory. Von Neumann has much earlier categorically expressed this view---``First of all we must admit that this objection [time being just a number] points at an essential weakness which is, in fact, the chief weakness of quantum mechanics. In fact, while all other quantities are represented by operators, there corresponds to time an ordinary number-parameter $t$, just as in classical mechanics.''\cite{nuemann1}

But how did {\it time} acquire its notorious parametric label? If time was an observable represented by a self-adjoint operator, then this operator would be canonically conjugate with the Hamiltonian, in accordance with quantum dynamics. But as early as 1933, Pauli\cite{pauli} has ``shown'' that the existence of a self-adjoint time operator canonically conjugate with a Hamiltonian implies that the time operator and the Hamiltonian have completely continuous spectra spanning the entire real line, a result widely known as Pauli's theorem. Thus for a semibounded or discrete Hamiltonian, no self-adjoint time operator would exist. Since physical systems are assumed to have a stable ground state, the Hamiltonian will generally be semibounded. Pauli then concluded that any ``attempt of introducing time as an operator in quantum mechanics must be fundamentally abandoned, and that the time $t$ in quantum mechanics has to be regarded as an ordinary number." From then on, it has been believed that no self-adjoint time operator exists in quantum mechanics,\cite{sakurai,pauli,busch1,busch5,delgado,giannitrapani,gottfried,holevo1,holevo2,cohen,park,srinivas,olhovsky,toller2} and has been widely recognized that time is a parameter merely serving to mark the evolution of a quantum system.\cite{sakurai,omnes}

However, it is likewise widely recognized that time undoubtedly acquires dynamical significance in questions involving the occurrence of an event,\cite{busch1,srinivas,msi} e.g. when a nucleon decays,\cite{eisenberg} or when a particle arrives at a given spatial point,\cite{muga,grot} or when a particle emerges from a potential barrier.\cite{landauaer}  Moreover, there is the time-energy uncertainty principle begging for an interpretation, a reasonable interpretation of which requires more than a parametric treatment of time.\cite{busch1,aharonov,bauer,busch2,fick2,fock,heisenberg,hilgevoord,hilgevoord1} It has then become part of the physicist's common sense that if one acknowledges the legitimacy of these problems and attempts to find quantum mechanical solutions to them, then one must solve them without contradicting Pauli's theorem. One recourse is to abandon the standard framework of quantum mechanics, and build a framework that can  support the introduction of time operators or that can accommodate the temporal aspects of quantum mechanics.\cite{eisenberg,holland,halliwell,blanchard} 

Another recourse is to stick with the single Hilbert space formulation; but it is acknowledged that this cannot be done without a compromise: {\it If one imposes self-adjointness as a desirable requirement for any observable, then one necessarily has to abandon the requirement that time operator be conjugate to the Hamiltonian; if, on the contrary, one decides that any proper time operator must be strictly conjugate to the Hamiltonian, then one has to renounce the search for a self-adjoint operator}.\cite{atkitson} In recent years, the problem of introducing time in quantum mechanics has taken the later route, abandoning self-adjointness in favor of the canonical commutation relation. And this specifically calls for extending quantum observables to maximally symmetric but not necessarily self-adjoint operators; in which case, quantum observables are generally positive operator valued measures.\cite{busch5,holevo1,holevo2}

However, we have shown recently that Pauli's theorem does not hold in single Hilbert spaces, and thus there is no a priori reason to exclude the existence of self-adjoint operators canonically conjugate to a  semibounded Hamiltonian, contrary to the claim of Pauli.\cite{galapon} Moreover, we have explicitly proved that to every discrete, semibounded Hamiltonian with compact inverse there exists a characteristic self-adjoint time operator conjugate to the Hamiltonian in a dense subspace of the system Hilbert space.\cite{galapon1} Furthermore, we have likewise shown that the non-self-adjoint free time of arrival operator in unbounded space, generally considered as an explicit demonstration of Pauli's theorem, defines a class of bounded, self-adjoint, and canonical time operator for a spatially confined particle.\cite{galapon2} 

These results inevitably force us to reassess our opinions on the role of time in single Hilbert space quantum mechanics and what constitutes a quantum time operator. To this end, we tackle on this paper two related issues that have been grossly neglected while Pauli's theorem was in force. First, we point out the existence of a large class of solutions to the canonical commutation relation in a Hilbert space, and argue that each solution can be identified as solution to a specific physical problem, say, to one facet of the quantum time problem. Second, we demonstrate the existence of multiple self-adjoint time operator solutions to the time-energy canonical commutation relation for a given discrete, semibounded Hamiltonian. Our discussion is focused on the interplay between these two in explicating self-adjoint time operators in single Hilbert space quantum mechanics.

The paper is then organized as follows. In Section-\ref{pau} we revisit Pauli's theorem. In Section-\ref{cano} we discuss the basic properties of the Hilbert space solutions to the canonical commutation relation (CCR). In Section-\ref{ans} we address the question on the physical relevance of the different solutions to the CCR. In Section-\ref{exam} we illustrate that there may be more than one self-adjoint time operators corresponding to a given Hamiltonian. Finally, in Section-\ref{what} we synthesize our discussion on the previous two sections, and address the issue of POVM-time observables.

\section{Pauli's Theorem: It's traditional and modern readings}\label{pau}

Pauli's argument goes as follows. Assume that there exists a self-adjoint time operator $T$, i.e an operator canonically conjugate with the Hamiltonian $H$, $\left[T,H\right]=iI$. Since $T$ is self-adjoint, the operator $U_\epsilon=\exp(-i\epsilon T)$ is unitary for all real number $\epsilon$.  Now let $\varphi_E$ be an eigenvector of $H$ with the eigenvalue $E$, then, according to Pauli, we have the implication 
\begin{equation}\label{i1}
[T,H]=i I \longrightarrow U_{\epsilon}\varphi_E=\varphi_{E+\epsilon},
\end{equation}
which further implies that $H$ has a continuous spectrum spanning the entire real line because $\epsilon$ is an arbitrary real number. Hence, Pauli concluded, if the Hamiltonian is semibounded or discrete, no self-adjoint time operator $T$ will exist to satisfy $\left[T,H\right]=iI$, otherwise, the operator $U_{\epsilon}$ will map the discrete or semibounded spectrum of $H$ into the entire real line, which is not possible for unitary $U_{\epsilon}$. 

A modern interpretation of the theorem is that if $P_E$ is the spectral decomposition of the (self-adjoint) Hamiltonian, then we have the implication
\begin{equation}\label{i2}
[T,H]=i I \longrightarrow U_{\epsilon}P_E U_{\beta}^{\dagger}=P_{E+\epsilon}.
\end{equation}
By reversing the roles of the Hamiltonian and the time operator in Pauli's argument, one gets similar conclusion about the properties of the time operator. Specifically, if a self-adjoint time operator exists such that its spectral decomposition is $P_T$, then we have the similar implication
\begin{equation}\label{i3}
[T,H]=i I \longrightarrow V_{t}P_E V_{t}^{\dagger}=P_{T+t},
\end{equation}
where $V_t=\exp(-itH)$, for all real $t$. The right-hand sides of the last two alleged implications identify the Hamiltonian and the time operator pair as forming a system of imprimitivities over the real line. Of course it is well-known that a system of such pair has the property that the operators have continuous spectra taking values in the entire real line. If one upholds the validity of the above implications, then one recovers Pauli's original conclusion for semibounded or discrete Hamiltonians---no self-adjoint time operator exists. 

These readings of Pauli's theorem imply that self-adjointness and canonicality of a time operator can not be imposed simultaneously: If canonicality is required, self-adjointness has to be renounced; on the other hand, if self-adjointness is required, canonicality has to be renounced. Moreover, they give the impression that any time operator is not only canonically conjugate with the Hamiltonian but {\it must} also be a generator of energy shifts.

However, we have recently shown and demonstrated that Implication-\ref{i1} does not hold in Hilbert space, and have argued that there are no Implications-2 and 3.\cite{galapon} These explicitly belie the belief that an operator canonically conjugate with a semibounded Hamiltonian is necessarily a generator of energy shifts and can not be self-adjoint. They inevitably put into question the traditional reading of what is a time operator in quantum mechanics. If there were no Implications-1, 2, and 3, then is there any justification in requiring the time operator and Hamiltonian pair to form a system of imprimitivies, as has been since Pauli, or is it just enough to require that time operators be canonically conjugate with the Hamiltonian? 

\section{Canonical Pairs in Hilbert Spaces}\label{cano}

We cannot answer the above questions without a clear understanding of the properties of a canonical pair in a Hilbert space. It has been the lack of understanding of these properties that led many to numerous false conclusions and unwarranted generalizations concerning the existence and non-existence of self-adjoint time operators. To the physicist, a canonical pair is a pair of operators $(Q,P)$ satisfying the canonical commutation relation, $[{Q},{P}]=i\opr{I}$, (CCR). Much of the inferred properties of $Q$ and $P$ have been derived from formal manipulations and have been assumed to hold in Hilbert spaces. Unfortunately, these inferences are generally valid only under some strict, unstated  conditions, which may exclude the assumed range of validity of the inferred properties. So the first step to a better perspective on the quantum time problem is to understand the properties of the CCR in Hilbert spaces. We stick with the basics.

If we seek a pair of Hilbert space operators, $Q$ and $P$,\footnote{A Hilbert space operator $A$ is more accurately denoted by $\oprD{A}$, where $\dom{A}$ is the domain of $A$ in $\mcal{H}$.} satisfying the CCR, then two things must be borne in  mind. First, no pair $(Q,P)$ exists to satisfy the CCR in the entire $\mcal{H}$; that is, there are no $Q$ and $P$ such that $[Q,P]{\varphi}=i{\varphi}$ for all ${\varphi}$ in $\mcal{H}$, or $[Q,P]=iI_{\mcal{H}}$, where $I_{\mcal{H}}$ is the identity in $\mcal{H}$. Had such a pair existed, the operators $Q$ and $P$ would have to be everywhere defined or bounded; and the CCR would lead to $[Q,P^n]=inP^{n-1}$, which, upon taking the norm of both sides, yields the inequality $2\norm{Q}\norm{P}\geq n$ for all $n>1$, which is a contradiction for bounded $Q$ and $P$, and for arbitrarily large integer $n$. This is Weidlant's theorem. A pair $(Q,P)$ then can at most satisfy the CCR in a---proper---subspace, $\mcal{D}_c$, of $\mcal{H}$; that is, the relation $[Q,P]{\varphi}=i{\varphi}$ holds only for all those ${\varphi}$ in $\mcal{D}_c$, where $\mcal{D}_c$ is always smaller than $\mcal{H}$. Thus a canonical-pair in a Hilbert space is a triple $\mcal{C}(Q,P; \mcal{D}_c)$---a pair of Hilbert space operators, $Q$ and $P$, together with a non-trivial, proper subspace $\mcal{D}_c$ of $\mcal{H}$, which we shall hereafter refer to as the canonical domain.

Second, there are canonical pairs in the {\it same} Hilbert space that are not unitarily equivalent, or pairs with distinct properties, say, spectral properties. This can be best appreciated by giving an example. Let $\mcal{H}=L^2(-\infty,\infty)$, the Hilbert space of square integrable complex valued functions in the real line. The pair of operators
\begin{displaymath}
(Q_1\varphi)\!(q)=q\varphi(q),\;\; (P_1\varphi)\!(q)=-i\varphi'(q),\nonumber
\end{displaymath}
together with the dense subspace $\mcal{D}_c^{1}\!\subset\!\mcal{H}$, consisting of all infinitely differentiable complex valued functions with compact support, forms the canonical pair $\mcal{C}_1\!(Q_1,P_1;\mcal{D}_c^{1})$; moreover, $Q_1$ and $P_1$ are essentially self-adjoint in $\mcal{D}_c^1$. Also the pair of operators\cite{fuglede}
\begin{displaymath}
(Q_2\varphi)\!(q)=q\varphi(q)+\varphi\!(q+i\sqrt{2\pi}),\;\;\; (P_2\varphi)\!(q)=-i\varphi'(q)+e^{-\sqrt{2\pi}q}\varphi(q),\nonumber
\end{displaymath}
together with the dense subspace $\mcal{D}_c^2\!\subset\!\mcal{H}$, consisting of the linear span of $q^n\, e^{-rq^2+cq}$, with $n=0,1,2,\dots$, $r>0$, and $c$ a complex number, forms the canonical pair  $\mcal{C}_2\!(Q_2,P_2;\mcal{D}_c^{2})$; moreover, $Q_2$ and $P_2$ are likewise essentially self-adjoint in $\mcal{D}_c^2$. Now the pairs $\mcal{C}_1$ and $\mcal{C}_2$ reside in the same Hilbert space $\mcal{H}$, yet they do not share the same properties. For one, the self-adjoint extensions of $Q_1$ and $P_1$, $\bar{Q}_1$ and $\bar{P}_1$, satisfy the Weyl relation $U(s)V(t)=e^{ist} V(t)U(s)$, for all real numbers $s$ and $t$, where $V(t)=\exp(it \bar{Q}_1)$ and $U(s)=\exp(is\bar{P}_1)$. On the other hand, the self-adjoint extensions of $Q_2$ and $P_2$ do not satisfy the same relation. This means that $\mcal{C}_1$ and $\mcal{C}_2$ are two distinct canonical pairs in $\mcal{H}$.

Clearly there could be numerous distinct canonical pairs in a given Hilbert space $\mcal{H}$. We shall refer to each pair as a Hilbert space {\it solution}, or simply a solution, to the CCR. Generally solutions split into two major {\it categories}, according to whether the canonical domain $\mcal{D}_c$ is dense or closed. We will call a canonical pair of {\it dense-category} if the corresponding canonical domain is dense; otherwise, of {\it closed-category} if the corresponding canonical domain is closed. Solutions under these categories further split into distinct {\it classes} of unitary equivalent pairs, and each class will have each own set of properties. Under such categorization of solutions the CCR in a given Hilbert space $\mcal{H}$, assumes the form $[Q,P]\subset i \mcal{P}_c$, where $\mcal{P}_c$ is the projection operator onto the closure $\bar{\mcal{D}}_c$ of the canonical domain $\mcal{D}_c$. If the pair $\mcal{C}$ is of dense category, then the closure of $\mcal{D}_c$ is just the entire $\mcal{H}$, so that $\mcal{P}_c$ is the identity $I_{\mcal{H}}$ of $\mcal{H}$. One should immediately recognize that we are considering a more general solution set to the CCR than has been considered so far. The traditional reading of the CCR in $\mcal{H}$ is the form $[Q,P]\subset i I_{\mcal{H}}$, which is just the dense category.

The pair $\mcal{C}_1$ above and all its unitary equivalents are then canonical pairs of dense categories. These pairs satisfy the Weyl relation and are unbounded with completely continuous spectrum taking values in the entire real line. On the other hand, the pair $\mcal{C}_2$ and all its unitary equivalents are canonical pairs of dense categories as well, but they do not satisfy the Weyl relation.  These later pairs have different spectral properties from the former. Clearly, these sets of pairs belong to different classes. They are not unitarily equivalent, and they represent two distinct classes of solutions of dense categories to the canonical commutation relation. Later we will give an example of a canonical pair of closed category in relation to the quantum time problem.

A question immediately arises---Is there a preferred solution to the CCR? That is, should we accept only solutions of dense or closed category of a specific class?

\section{An answer from the Position-Momentum canonical pairs}\label{ans}

Let us refer to the well-known position and momentum operators in three different configuration spaces: The entire real line, $\Omega_1=(-\infty,\infty)$; the bounded segment of the real line, $\Omega_2=(0,1)$; and the half line $\Omega_3=(0,\infty)$. Quantum mechanics in each of these happens in the Hilbert spaces $\mcal{H}_1=L^2(\Omega_1)$,  $\mcal{H}_2=L^2(\Omega_2)$, $\mcal{H}_3=L^2(\Omega_3)$, respectively. The position operators, $\opr{Q}_j$,  in $\mcal{H}_j$, for all $j=1,2,3$, arise from the fundamental axiom of quantum mechanics that the propositions for the location of an elementary particle in different volume elements of $\Omega_j$ are compatible (see Jauch\cite{jauch} for a detailed discussion for $\Omega_1$, which can be extended to $\Omega_2$ and $\Omega_3$). They are self-adjoint and are given by the operators $(\opr{Q}_j\varphi)\!(q)=q\varphi(q)$ for all $\varphi$ in the domain $\dom{\opr{Q}_j}=\left\{\varphi\in\mcal{H}_j:\, \opr{Q}_j\varphi\in\mcal{H}_j\right\}$. Note that $\opr{Q}_1$ and $\opr{Q}_3$ are both unbounded, while $\opr{Q}_2$ is bounded.

Now each of the configuration spaces, $\Omega_1$, $\Omega_2$, and $\Omega_3$, has an identifying property. $\Omega_1$ is fundamentally homogeneous---points in there are physically indistinguishable. On the other hand, $\Omega_2$ and $\Omega_3$ are not homogeneous, the boundaries being the distinguishing factor. However, their inhomogeneities are not the same: $\Omega_2$ has two boundaries, while $\Omega_3$ has one. These properties can be expressed mathematically in terms of the respective representation of translation in each of these configuration spaces. Translation in $\Omega_1$ is isomorphic to the additive group of real numbers; in $\Omega_2$, to the group of rotations of the circle\footnote{The proper treatment of $\Omega_2$ is more elaborate than our treatment here. Our treatment is sufficient though for our present purposes.}; in $\Omega_3$, to the semigroup of additive positive numbers. Thus in $\mcal{H}_1$ and $\mcal{H}_2$ there are one parameter unitary operators $\mcal{U}_1(s)$, $\mcal{U}_2(s)$ representing translations in $\mcal{H}_1$ and $\mcal{H}_2$, respectively. And in $\mcal{H}_3$ there is a completely one-parameter semigroup $\mcal{U}_3(s)$ representing translations. 

The properties of the three configuration spaces can now be explicitly stated in the following respective forms
\begin{displaymath}
\Omega_1:\;\;\;\left(\mcal{U}_1^{\dagger}(s)\varphi\right)\!(q)=\varphi(q-s),\;\;\mbox{for all}\;\; s\in\Re,
\end{displaymath}
\begin{displaymath}
\Omega_2:\;\;\; \left(\mcal{U}_2^{\dagger}(s)\varphi\right)\!(q)= \left\{\begin{array}
                        {c@{\quad:\quad}l}
                        \varphi(q-s) & \mbox{for}\;\; 1>q>s>0\\
                        \varphi(1+(q-s)) & \mbox{for}\;\;1>s>q>0
                        \end{array}\right.,
\end{displaymath}
\begin{displaymath}
\Omega_3:\;\;\; \left(\mcal{U}_3^{\dagger}(s)\varphi\right)\!(q)= \left\{\begin{array}
                        {c@{\quad:\quad}l}
                        \varphi(q-s) & \mbox{for}\;\; q>s\\
                        0 & \mbox{for}\;\;q<s
                        \end{array}\right. .
\end{displaymath}
If we define the momentum operator as the generator of translation in the configuration space, then the momentum operator in $\mcal{H}_j$ is the operator $\opr{P}_j$ defined on all vectors $\varphi$ for which the limit $\lim_{s\to 0}(i s)^{-1}(\mcal{U}_j(s)-\mcal{I}_{\mcal{H}})\varphi=\opr{P}_j\varphi$ exists. Explicitly, it is given by $(\opr{P}_j\varphi)\!(q)=-i\varphi'(q)$.

In each $\mcal{H}_j$, there exists a dense common subspace $\mcal{D}_j$ of $\opr{Q}_j$ and $\opr{P}_j$, which is invariant under $\opr{Q}_j$ and $\opr{P}_j$, for which we have the canonical pair $\mcal{C}_j(\opr{Q}_j,\opr{P}_j;\mcal{D}_j)$. The $\mcal{C}_j$'s are of the same dense category, but they belong to different classes: $\opr{Q}_1$ and $\opr{P}_1$ are both self-adjoint, having absolutely continuous spectra spanning the entire real line $\Re$ and forming  a system of imprimitivities in $\Re$, and their restrictions in $\mcal{D}_1$ are essentially self-adjoint. $\opr{Q}_2$ is self-adjoint with an absolutely continuous spectra in $(a,b)$, and its restriction in $\mcal{D}_2$ is essentially self-adjoint; $\opr{P}_2$ is self-adjoint with a pure point spectrum, but its restriction in $\mcal{D}_2$ is not essentially self-adjoint. $\opr{Q}_3$ is self-adjoint with an absolutely continuous spectra in $(0,\infty)$, and its restriction in $\mcal{D}_3$ is essentially self-adjoint; $\opr{P}_3$ is maximally symmetric and non-self-adjoint, thus without any self-adjoint extension. These varied properties of the position and momentum canonical pairs are obviously the consequences of the underlying properties of their respective configuration spaces.

Now we can go back to the question we have posed in the previous section. {\it Is there a preferred solution to the CCR?} Recall that there is only one separable Hilbert space; that is, all separable Hilbert spaces are isomorphically equivalent to one other, so that there are unitary operations transforming one Hilbert space to another. The three Hilbert spaces, $\mcal{H}_1$, $\mcal{H}_2$, and $\mcal{H}_3$, are separable, and hence can be transformed to a common Hilbert space $\mcal{H}_C$, together with all the operators in them, including their respective position and momentum operators. The canonical pairs, $\{\mcal{C}_1$, $\mcal{C}_2$, $\mcal{C}_3\}$, are then solutions of the CCR in the same Hilbert space $\mcal{H}_C$. And we have seen that they are of dense category solutions, but of different classes. If we look at the diverse properties of the above $\mcal{C}_j$'s, we can see that these properties are reflections of the fundamental properties of the underlying configuration spaces. It is then misguided to prefer one solution of the CCR over the rest or to require a priori a particular category of a specific class of a solution without a proper consideration of the physical context against which the solution is sought. For example, if we insist that only canonical pairs forming a system of imprimitivities over the real line are acceptable, then, within the context of position-momentum pairs, we are insisting homogeneity in all configuration spaces, a preposterous proposition. Why impose the homogeneity of, say, $\Omega_1$ in intrinsically inhomogeneous configuration spaces like $\Omega_2$ and $\Omega_3$?

The example of the position and the momentum operators makes it clear that the set of properties of a specific solution to the CCR is consequent to a set of underlying fundamental properties of the system under consideration, or to the basic definitions of the operators involved, or to some fundamental axioms of the theory, or to some postulated properties of the physical universe. That is to say that a specific solution to the CCR is canonical in some {\it sense}, i.e. of a particular category and of a particular class. It is conceivable to impose that a given pair be canonical as a priori requirement based, say, from its classical counterpart, but not the sense the pair is canonical without a deeper insight  into the underlying properties of the system. In other words, we don't impose in what {\it sense} a pair is canonical if we don't know much, we derive in what {\it sense} instead. Furthermore, if a given pair is known to be canonical in some {\it sense}, then we can learn more about the system or the pair by studying the structure of the {\it sense} the pair is canonical.

\section{Self-adjoint time operators}\label{exam}

Pauli's theorem has made the impression that an operator canonically conjugate with a Hamiltonian, i.e. a time operator, is necessarily a generator of energy shifts or an operator of an imprimitivity pair in the real line, and thus cannot be self-adjoint. With Implications-\ref{i1}, 2 and 3 exhibited not to hold in Hilbert spaces, there thus exists no a priori reason for the non-existence of self-adjoint time operators for semibounded Hamiltonians. However, with the conjugacy of the time operator with the Hamiltonian and imprimitivity of the time-operator-Hamiltonian-pair taken synonymous for a long time, one now wonders what a self-adjoint time operator is without the imprimitivity requirement. It is in this context that the need for the appreciation of the different solutions to the CCR becomes necessary. In this section, we illustrate how self-adjoint time operators can arise without satisfying the imprimitivity requirement, and demonstrate that a given Hamiltonian can in fact form a canonical pair with two self-adjoint time operators of different categories, one dense and another closed.

Let us consider a particle confined between two points with length $2l$, subject to no force in between the boundaries. 
We attach the Hilbert space $\mcal{H}=L^2[-l,l]$ to the system. The position  operator is unique and is  given by the bounded operator $\opr{q}$,  $\left(\opr{q}\varphi\right)\!(q)=q\varphi(q)$ for all $\varphi$ in $\mcal{H}$. On the other hand, the momentum operator and the Hamiltonian are not unique, and have to be considered carefully. We assume the system to be conservative and we require that the evolution of the system be generated by a purely kinetic Hamiltonian. The former requires a self-adjoint Hamiltonian to ensure that time evolution is unitary. The latter requires a self-adjoint momentum operator commuting with the Hamiltonian. These requirements are only satisfied by the following choice of the momentum operator. For every $\abs{\gamma}<\pi$, define the self-adjoint momentum operator $(\opr{p_{\gamma}}\phi)\!(q)=-i \phi'(q)$, with domain $\dom{p_{\gamma}}$ consisting of those vectors $\phi$ in $\mcal{H}$ such that $\int\abs{\phi'(q)}^2\,dq<\infty$, and satisfying the boundary condition  $\phi(-l)=\expo{-2i\gamma}\phi(l)$. With $\opr{p}_{\gamma}$ self-adjoint, the Hamiltonian is purely kinetic, i.e. $\opr{H_{\gamma}}=(2\mu)^{-1}\opr{p}_{\gamma}^2$.  The momentum and the Hamiltonian then commute and have the common set of eigenvectors $\phi_k^{(\gamma)}(q)=(2l)^{-\frac{1}{2}} \exp\!\left(i\,(\gamma+k \pi) \frac{q}{l}\right)$, with respective eigenvalues $p_{k,\gamma}= (\gamma+k\pi)l^{-1}$, $E_k=p_{k,\gamma}^2(2\mu)^{-1}$, for all $k=0,\pm1,\pm2\cdots$. In the following, we give two time operators of different categories for the Hamiltonian $\opr{H}_{\gamma}$.

\subsection{A self-adjoint time operator of dense category}

There exists a compact and self-adjoint operator $\opr{T}_c^{\gamma}$ such that $\opr{T}_c^{\gamma}$ and $\opr{H}_{\gamma}$ form a canonical pair of dense category. This operator has the integral representation 
\begin{displaymath}
\left(\opr{T}_c^{\gamma}\varphi\right)\!(q)=\int_{-l}^{l}\kernel{\opr{T}_c^{\gamma}}\varphi(q')\,dq',
\end{displaymath}
whose kernel is given by
\begin{displaymath}
\kernel{\opr{T}_c^{\gamma}}=i\sum_{k,k'}\!'\frac{\varphi_k^{(\gamma)}(q) \varphi_{k'}^{(\gamma)}(q')^*}{E_k-E_{k'}},
\end{displaymath}
where the primed sum indicates that $k=k'$ is excluded from the summation. That is, the pair $\opr{T}_c^{\gamma}$ and $\mcal{H}_{\gamma}$ satisfy the canonical commutation relation in some dense subspace $\mcal{D}_c^1$ of $\mcal{H}$,
\begin{displaymath}
\left([\opr{T}_c^{\gamma},\opr{H}_{\gamma}]\varphi\right)\!(q)=i\varphi(q),\;\; \mbox{for all}\;\; \varphi(q)\in\mcal{D}_c^1
\end{displaymath}
\begin{displaymath}\label{domdom}
\mcal{D}_c^{1}=\left\{\varphi(q)=\sum_k a_k \varphi_k^{(\gamma)}(q),\; \sum_k \abs{a_k}^2<\infty,\; \sum_k a_k =0\right\}.
\end{displaymath}
Since the canonical domain is dense, i.e. orthogonal only to the zero vector, the canonical pair $\mcal{C}(\opr{T}_c^{\gamma},\opr{H}_{\gamma};\mcal{D}_c^1)$ is of dense category. While $\opr{T}_c^{\gamma}$ and $\opr{H}_{\gamma}$ form a canonical pair, they do not form a system of imprimitivies over the real line since $\opr{T}_c^{\gamma}$ is compact.

But what is $\opr{T}_{c}^{\gamma}$? In a separate work we have explicitly shown that to every discrete, semibounded Hamiltonian with constant degeneracy and with compact inverse there exists a time operator characteristic of the Hamiltonian.\cite{galapon1} The Hamiltonian $\opr{H}_{\gamma}$ satisfies all these conditions and the operator $\opr{T}_{c}^{\gamma}$ is the corresponding characteristic time operator to $\opr{H}_{\gamma}$. But what is the physical content of the characteristic time operator? It is sufficient at this moment to say that its any two dimensional projection can serve as a quantum clock. Let us consider the general case. Given $k$ and $l$, consider the closed subspace spanned by the eigenstates $\varphi_k$ and $\varphi_l$ of the Hamiltonian. Denote this subspace by $\mcal{H}_{kl}$. Obviously the state $\varphi_{kl}=2^{-\frac{1}{2}}(\varphi_k-\varphi_l)$ in $\mcal{D}_c$ belongs to $\mcal{H}_{kl}$. Let $\opr{P}_{kl}$ be the projection operator onto $\mcal{H}_{kl}$. Let $\opr{T}_{kl}=\opr{P}_{kl}\opr{T}\opr{P}_{kl}=i\omega_{kl}^{-1}\left(\ketbra{\varphi_k}{\varphi_l}-\ketbra{\varphi_l}{\varphi_k}\right)$, where $\omega_{kl}=(E_k-E_l)$; and let $\opr{H}_{kl}=\opr{P}_{kl}\opr{H}\opr{P}_{kl}=E_k\ketbra{\varphi_k}{\varphi_k}+E_l\ketbra{\varphi_l}{\varphi_l}$.
Since $\opr{T}$ and $\opr{H}$ are canonically conjugate in $\mcal{D}_c$, we expect that $\opr{T}_{kl}$ and $\opr{H}_{kl}$ continue to satisfy the canonical commutation relation in the subspace of $\mcal{H}_{kl}$ spanned by $\varphi_{kl}$, i.e. $\left(\opr{T}_{kl}\opr{H}_{kl}-\opr{H}_{kl}\opr{T}_{kl}\right)\varphi_{kl}=i\varphi_{kl}$. The state $\varphi_{kl}$ evolves according to $\varphi_{kl}(t)=2^{-\frac{1}{2}}\left(e^{-i E_k t} \varphi_k-e^{-i E_l t} \varphi_l\right)$. Now $\opr{T}_{kl}$ has the expectation value and variance
\begin{displaymath}\label{tat}
\braket{\varphi_{kl}(t)}{\opr{T}_{kl}\varphi_{kl}(t)}=\frac{1}{\omega}\,\mbox{Sin}\, \omega_{kl} t,\;\;\;
\Delta T_{\varphi_{kl}(t)} \, \Delta H_{\varphi_{kl}(t)} = \abs{\mbox{Cos}\, \omega_{kl} t} \frac{1}{2}.
\end{displaymath}
Here we see that the observable $\opr{T}_{kl}$ and the entire $\mcal{D}_c$ wrap the entire time axis into the circle. Thus for a given $t$ there is a positive integer $k$ and a time interval $\tau$ such that $t=\tau+2\pi n \,\omega_{kl}$. For a given $n$ and for small $\tau$'s, equations (\ref{tat}) reduce to
\begin{displaymath}
\braket{\varphi_{kl}(t(\tau))}{\opr{T}_{kl}\varphi_{kl}(t(\tau))}= \tau, \;\;
\Delta T_{\varphi_{kl}(t(\tau))} \, \Delta H_{\varphi_{kl}(t(\tau))} = \frac{1}{2}.
\end{displaymath}
The operator $\opr{T}$ then is a quantum clock wrapping the entire time axis into the circle and saturating the time energy uncertainty relation in every neighborhood of $\abs{t-2\pi n\omega_{kl}}$. We shall give a more detailed analysis of characteristic time operators elsewhere.

\subsection{A self-adjoint time operator of closed category}

There exists a compact and self-adjoint operator $\opr{T}^{\gamma}$, $\gamma\neq 0$, such that $\opr{T}^{\gamma}$ and $\opr{H}_{\gamma}$ form a canonical pair of closed category. This operator has the integral representation 
\begin{displaymath}
\left(\opr{T}^{\gamma}\varphi\right)\!(q)=\int_{-l}^{l}\kernel{\opr{T}^{\gamma}}\varphi(q')\,dq',
\end{displaymath}
whose kernel is given by
\begin{displaymath} \label{repre}
	\kernel{\opr{T}^{\gamma}}=\frac{\mu}{4\,\sin\!\!\gamma} (q+q')\left(e^{i\,\gamma} 				\,\mbox{H}(q-q')+e^{-i\,\gamma}\, \mbox{H}(q'-q) \right)
\end{displaymath}
where H$(q-q')$ is the Heaviside function. That is, the pair $\opr{H}_{\gamma}$ and $\opr{T}^{\gamma}$ satisfies the canonical commutation relation in a closed subspace of $\mcal{H}$,
\begin{equation}\label{close}
	\left([\opr{T}^{\gamma},\opr{H_{\gamma}}]\varphi\right)\!\!(q) =i\,\varphi(q)\;\;\;\;\mbox{for all}\;\;\varphi(q)\in	\mcal{D}_{c}^{\gamma}
\end{equation}
\begin{displaymath}
 \mcal{D}_c^{\gamma}=\left\{\int_{-l}^{l} \varphi(q)\,dq=0,\, \varphi(\partial)=0,\,
				\varphi'(\partial)=0 \right\}.
\end{displaymath}
Since the canonical domain $\mcal{D}_{c}^{\gamma}$ is closed, i.e. orthogonal to the one-dimensional subspace $\mcal{D}^{(\gamma) \perp}=\left\{\phi=c, \, c\in \mcal{C}\right\}$ the canonical pair $\mcal{C}(\opr{T}^{\gamma},\opr{H}_{\gamma};\mcal{D}_c^{\gamma})$ is of closed category. Also, while $\opr{T}^{\gamma}$ and $\opr{H}_{\gamma}$ form a canonical pair, they do not form a system of imprimitivies over the real line since $\opr{T}^{\gamma}$ is compact.

But what is $\opr{T}^{\gamma}$? This operator is the quantization of the classical passage time $T=\mu q/p$ in the Hilbert space $\mcal{H}=L^2(-l,l)$, subject to the requirement that the quantum Hamiltonian is the quantization of the purely kinetic classical Hamiltonian of the freely evolving particle between the boundaries. This condition leads to the momentum operator $\opr{p}_{\gamma}$ and the Hamiltonian $\opr{H}_{\gamma}$ given above. For a given $\gamma\neq 0$, we have the following correspondences 
\begin{eqnarray}
\mcal{Q}:H=\frac{p^2}{2\mu}\;\;\;& \mapsto & \;\;\;\opr{H}_{\gamma}=\frac{\opr{p}_{\gamma}^2}{2\mu},\nonumber\\
\mcal{Q}:T=\mu \frac{q}{p}\;\;\; & \mapsto & \;\;\;\opr{T}^{\gamma}=\frac{\mu}{2}\left(\opr{q}\opr{p}_{\gamma}^{-1}+\opr{p}_{\gamma}^{-1}\opr{q}\right),\nonumber\\
\mcal{Q}:\left\{T,H\right\}=1\;\;\; & \mapsto & \;\;\;\left[\opr{T}^{\gamma},\opr{H}_{\gamma}\right]\,\subset \,i\, \mcal{I}_{\gamma},\label{3rd}
\end{eqnarray}
where $\mcal{Q}$ is a quantization and $\mcal{I}_{\gamma}$ is the identity in the closure of the canonical domain $\mcal{D}_c^{\gamma}$. Notice that relation-\ref{3rd}, the right hand side of which is just equation-\ref{close}, is Dirac's Poisson-bracket-commutator correspondence at work. It is interesting to note though that Dirac's correspondence principle holds in a closed subspace for the $(\opr{T}^{\gamma},\opr{H}_{\gamma})$ pair, not in a dense subspace, as expected in the theory of quantization. Now $\opr{T}^{\gamma}$ is best appreciated by observing that the operator $\opr{T}_{\gamma}=-\opr{T}^{\gamma}$ is identifiable as the time of arrival operator at the origin for the spatially confined particle. We have referred to $\opr{T}_{\gamma}$ as the confined, non-periodic time of arrival operator for a given $\abs{\gamma}<\pi$.\cite{galapon2} $\opr{T}_{\gamma}$ possesses the expected set of symmetry of a time of arrival operator and its eigenfunctions are identifiably time of arrival states. That is positive eigenvalue eigenfunctions evolve to symmetrically collapse at the origin, while negative eigenvalue eigenfunctions evolve to symmetrically collapse at the origin in the time reversed direction. We refer the reader to our earlier work\cite{galapon2} for a fuller account of the confined time of arrival operators.

\section{Discussion}\label{what}

For a long time, Pauli's theorem has led most, if not all, to believe that the canonical commutation relation $[T,H]=iI$ only admits Hilbert space solutions that form a system of imprimitivities on the real line, contrary to the fact that there is a large class of solutions to the CCR, as we have discussed above. Consequently the theorem has brushed aside and downplayed the rest of the solution set of the canonical commutation relation for a given Hamiltonian. Our example clearly demonstrates that there are self-adjoint time operator solutions to $[T,H]=iI$ for semibounded Hamiltonians. It further demonstrates that for a given Hamiltonian there are possibly more than one class of solutions to the canonical commutation relation. In our example, the Hamiltonian is conjugate with two self-adjoint time operators belonging to two different categories. That these operators belong to distinct categories can be traced from the fact that they have distinct physical origins. For the time operator forming a dense-category with the Hamiltonian, it is characteristic of the system---it can be taken for an inherent quantum clock. For the time operator forming a closed category with the Hamiltonian, it is problem-specific---it is a direct result of quantization of the classical first passage time. As we have asserted in Section-\ref{ans}, a specific solution to the CCR is canonical in some {\it sense}, i.e. of a particular category and of a particular class, and its {\it sense} is consequent to a set of underlying fundamental properties of the system under consideration, or to the basic definitions of the operators involved. Our example clearly illustrates this assertion. Obviously the condition of imprimitivity is not necessary for a time operator to satisfy in order to be physically meaningful. 

But how about the current prevailing claim that time operators---in order to be meaningful---must be characterized as positive-operator-valued-measure (POVM) observables that transform covariantly under time translations?\cite{busch1,busch5,giannitrapani,srinivas,toller2,busch2,harald,egusquiza} Such a claim would require non-self-adjoint time operators for semibounded Hamiltonians. However, this claim has been introduced under the understanding that Pauli's theorem is the statement that {\it an operator canonically conjugate to a Hamiltonian is a generator of energy shifts}. But now we know that this traditional reading of Pauli's theorem, together with its modern rendering, is not correct in single Hilbert spaces. Should we then rule these covariant non-self-adjoint time operators misplaced?

It is misplaced to require covariance if one does so under the assumption that a time operator must {\it necessarily} be a generator of energy shifts. However, if one upholds the legitimacy of POVM to accommodate non-self-adjoint observables, as one should, then one may still be justified in requiring covariance as long as the required covariance is anchored on physical grounds. One must only acknowledge that covariant non-self-adjoint time operators are just one class and not the only class of solutions to the canonical commutation relation for a given Hamiltonian.  Covariance can then be seen as a specific property of one class of solutions to the canonical commutation relation that can be anchored on specific problems. One must note though that over requiring covariance can lead to non-normalizable positive operator valued measures\cite{srinivas} which results to missing probabilities. On this instance, a solution to the canonical commutation relation is in conflict with the axioms of quantum mechanics. And thought is required to consider whether they are acceptable or not, an acceptance of which requires further revision of the axioms of quantum mechanics.

\end{document}